# Exciton-photon complexes and dynamics in the concurrent strong-weak coupling regime of singular site-controlled cavity quantum electrodynamics

Jiahui Huang[1,†], Wei Liu[1,†,*], Murat Can Sarihan[1], Xiang Cheng[1], Alessio Miranda[2], Benjamin Dwir[2], Alok Rudra[2], Eli Kapon[2], Chee Wei Wong[1,*]

[1] Mesoscopic Optics and Quantum Electronics Laboratory, University of California, Los Angeles, CA 90095, USA

[2] Institute of Physics, École Polytechnique Fédérale de Lausanne, Lausanne, Switzerland

* Correspondence: weiliu01@ucla.edu; cheewei.wong@ucla.edu

[†]These authors contributed equally.

We investigate the exciton complexes photoluminescence, dynamics and photon statistics in the concurrent strong-weak coupling regime in our unique site-controlled singular inverted pyramidal InGaAs/GaAs quantum dots–photonic crystal cavities platform. Different from a clear boundary between strong and weak QD-cavity coupling, we demonstrate the strong and weak coupling can coexist dynamically, as a form of intermediate regime mediated by phonon scattering. The detuning-dependent micro-photoluminescence spectrum reveals concurrence of exciton–cavity polariton mode avoided crossing, as a signature of Rabi doublet of the strong coupled system, the blue shifting of coupled exciton–cavity mode energy near zero-detuning ascribed to the formation of collective states mediated by phonon assisted coupling, and their partial out-of-synchronization linewidth-narrowing linked to their mixed behavior. By detailing the optical features of strongly-confined exciton-photon complexes and the quantum statistics of coupled cavity photons, we reveal the dynamics and anti-bunching/bunching photon statistical signatures of the concurrent strong-weak intermediate coupled system at near zero-detuning. This study suggests our device has potential for new and subtle cavity quantum electrodynamical phenomena, cavity-enhanced indistinguishable single photon generation, and cluster state generation via the exciton-photon complexes for quantum networks.

Solid-state quantum-confined light-matter interactions serves as a critical resource for unbreakable secure quantum communications [1-5] and long-distance quantum communications through a trusted repeater node architecture [6-9], guided by rapid demonstrations of sub-Poissonian single photon sources [10, 11], strong coupling in high quality factor ($Q$) cavities and localized modes [12-17], precise positioning of single quantum dots (QDs) with nanometer-scale accuracy [18], entanglement generation [7, 19-22], and coherent control [10, 15, 23, 24]. Moreover, recent study on the quantum network protocol based on deterministic cluster state generation [22, 25-27] and solid-state quantum memory enabled single-photon switch [9] received considerable interests. In particular, cavity-enhanced biexciton-dark-exciton cascaded system in GaAs-based QD [28, 29] has always served as a promising platform to realize such protocol. Crucially, pure dephasing of two-level system, defined as any disruption of quantum state without causing population relaxation but introducing random phase evolution, plays a fundamental role in the error-tolerant network [30] and frequency-stabilized on-chip scalable indistinguishable single photon emitter [31, 32]. In the GaAs-based QDs, the main pure dephasing mechanisms, thereby their spectral broadening, are related to rapidly fluctuating electrical charges [33] and carrier-phonon interactions [13, 34, 35]. In the weak coupling regime, pure dephasing mediated Purcell effect assists a large fraction of the QD emission being channeled into the cavity mode, leading to an efficient off-resonance cavity feeding up to a few meV detuning [33, 36, 37] At the same time, finite coupling between QD exciton and cavity can also manifest the emission of cavity photon at the QD exciton energy such as in a dissipative cavity where cavity loss photon at a rate much faster than QD [36, 38]. Alternatively, in the strong coupling regime, the vacuum Rabi splitting and polariton state is modified by the damped coherent Rabi oscillations induced by pure dephasing [39]. Apart from the clear boundary between strong and weak coupling, recent studies indicate the possibility of a unique *intermediate* coupling regime, which shares the coexisting features of both strong and weak coupling in the self-assembled GaAs-based QDs in micropillar [40, 41] and photonic crystal (PhC) cavities [42]. Specifically, recent theoretical studies suggests phonon-assisted QD-cavity interaction play an important role in the intermediate coupling regime [43, 44]. It suggests the importance of understanding pure dephasing in a cavity quantum electrodynamical (cQED) system for both fundamental science and practical quantum applications.

Particularly, the intrinsic pure dephasing of two-level system in QDs and its interaction with cavity remain to be difficult to access. It is because most of the current studies rely on the self-assembled Stranski-Krastanov (SK) QDs, which induces spurious cavity feedings related to wetting layer [45]. We note that the hybridization of localized SK QD states with delocalized wetting layer states results in a broadband quasi-continuum of states that give rise to a background emission [46, 47]. In addition, the random SK QDs nucleation leads to unfavorable presence of parasitic QDs in the vicinity and brings ambiguity of the exact location of the presumed single QD inside the cavity, which results from misalignment of QD and cavity electric field, preventing clear identification of pure dephasing effects. Alternatively, without introducing wetting layer, nanoscale site-controlled high symmetric pyramidal QD grown on GaAs substrates patterned with inverted pyramidal recesses ensures deterministic single QD-cavity overlapping and strong suppression of the interaction between confined excitons with external environments [48, 49], an ideal platform for uncovering the complex cQED in concurrent strong-weak intermediate coupling regime. In this study, based on our unique singular site-controlled pyramidal InGaAs/GaAs QD – photonic crystal (PhC) cavities hybrid structure, we systematically investigate the intermediate coupling regime and related carrier recombination dynamics, using multiple spectroscopic modalities including high-resolution polarized micro-photoluminescence (µPL) time-resolved PL, and second-order photon correlation measurements. Moreover, we reveal insights into the optical features of exciton-photon complexes strongly confined in our site-controlled QD and the bunching/anti-bunching and collective state photon statistics of coupled exciton-cavity system, towards cavity-enhanced multi-partite quantum information processing and cluster state quantum network implementations.

Our singular site-controlled QD-cavity system is studied via high-resolution micro-photoluminescence (µPL). The sample is mounted on the cold finger of a liquid helium flow cryostat, positioned accurately with piezoelectric actuators in *xy*-direction. The sample is excited by either a continuous wave (CW) low-noise diode-pumped solid-state (DPSS) laser at 532 nm, or a 900 nm pulsed diode laser with ≈ 100 ps pulse duration. A 100× microscope objective with numerical aperture of 0.7 allows an ≈ 1 µm diffraction-limited pump beam spot size on the sample. The µPL signals are collected and collimated by the same objective and refocused onto the entrance slit of a 1-meter spectrometer with 1200 grooves/cm, enabling a high spectral resolution of 8 pm (10 µeV). At one port of the spectrometer dual exit, a liquid nitrogen-cooled charge-coupled device (CCD) acquires the PL spectrum. The other output is attached to a free-space Hanbury-Brown and Twiss (HBT) setup comprising of a 50:50 non-polarizing beamsplitter and two aligned Si avalanche single-photon detectors (APDs) with 25-Hz dark counts. Time-correlated single-photon counting (TCSPC) is implemented for the second-order correlation ($g^2$) and time-resolved PL (TRPL) measurements. Polarization-resolved µPL is performed by implementing a half-wave plate and a linear polarizer in front of spectrometer entrance. The linear polarizer is fixed at the polarization orientation in which the grating spectrometer has the maximum reflectivity. The measurement setup schematic, description of sample structure, and modeling of PhC cavity are detailed in Supplementary Section I.

**Pure dephasing and exciton-photon complexes in the singular site-controlled InGaAs QD-PhC cavity**

Figure 1a presents the resulting sharp excitonic emissions from a single pyramidal InGaAs QD and the L3-PhC cavity mode (CM) at 25 K with X-CM detuning ≈ 2 meV, excited non-resonantly by the 532 nm laser. As labeled in Figure 1a, the *s*-state excitonic emission [50, 51] consists of a complex of negatively-charged exciton ($X^-$), neutral exciton (X), biexciton (BX), and likely an excited light hole (LH) state. The relative binding energies [52] of the BX and $X^-$ with respect to X is -1.06 meV and 3.41 meV respectively. The pronounced $X^-$ population even at low excitation of 10 nW is mainly due to background-doping donors, incorporated during growth. Moreover, as a contribution factor, the significant asymmetric mobility between electron and hole further favors the formation of $X^-$, rather than the positively charged exciton ($X^+$) [53]. At the low excitation regime (<150 nW), the full-width half-maximum (FWHM) linewidths of the $X^-$, X, BX and LH transitions are nearly constant with $\gamma_{X-}$ = 173 ± 9 μeV, $\gamma_X$ = 143 ± 10 μeV, $\gamma_{XX}$ = 117 ± 42 μeV, and $\gamma_{LH}$ = 202 ± 5 μeV respectively, extracted by Lorentzian fitting. It is apparent that the linewidth broadening is larger than the Fourier transform limits of the typical QD nanosecond lifetime (≈ 7 μeV), suggesting the contribution of pure dephasing via fluctuating environmental charges. Though phonon scattering can contribute to the broadening, it is not expected to be dominant below 50 K [54]. The larger linewidth observed at low excitation for $X^-$ compared to X can be understood by its additional charge carriers which has increased Coulomb interactions with fluctuating environmental electric fields. We note that $\gamma_{X-}$ monolithically decreases by ≈ 30 μeV as excitation increases from 200 to 700 nW, which is due to the saturation of the local charge states. Meanwhile, the significantly larger $\gamma_{LH}$ is caused by a more delocalized LH state, experiencing stronger surrounding charge fluctuations. Apart from the excitonic emission, the CM line has a FWHM of $\kappa_{CM}$ of 296 ± 12 μeV (*Q* factor ≈ 4,230), which is in the dissipative cavity regime of $\kappa_{CM} \gg \gamma_{exciton}$, where $\kappa_{CM}$ and $\gamma_{exciton}$ are the cavity loss and QD radiative decay rates respectively.

It is worth pointing out that the absence of a 2D wetting layer in our single pyramidal QD growth rules out far-off-resonance cavity feeding by a spurious emission background. Furthermore, though the pyramidal QD nucleates in the vicinity of three 1-dimensional ridge quantum wires at the wedges of the pyramid, the quantum wire influence is shown to be negligible for low QD excitation powers, owing to the large energy difference and absence of hybridization between the localized QD states and 1D delocalized state and the lower mobility of charges in the disordered 1D barrier. To further verify the above assignment of each sharp transition line, Figure 1b presents the pump power-dependent µPL at 25 K. The pump power-dependent integrated µPL intensity of X, $X^-$ and LH in logarithmic scale [51] reveals a sublinear slope of $0.6 \pm 0.1$, $0.8 \pm 0.1$, and $0.6 \pm 0.1$ respectively, correlated with the CM increase. In contrast, the far-off resonance BX has the distinctive $2.0 \pm 0.2$ slope and features a saturation pump power larger than other excitonic emissions. In addition, the off-resonance CM as a function of pump power shows a slope of $1.1 \pm 0.1$.

By further increasing the pump power, ultrasharp lines appear next to the BX and X transitions (labeled with arrows in the magnified Figure 1c), which are likely due to the charged exciton-photon complexes and possible recombination of ground state QD electron with higher-order QD hole states, which have small (< meV) energy spacings. Particularly the sharp emission slightly below the BX line exhibits a super-linear power dependent behavior, which suggests the feature of a negatively-charged biexciton ($BX^-$) [55] or excited biexcitonic states [56]. Meanwhile, the rich lines ≈ 1 meV blue-shifted from BX resembles the positively-charged complexes [57]. This can be attributed to the fact that the QD hole capture rate is proportional to the excitation power [58]: the increased direct hole captured by the QD at stronger pump excitation leads to a population from negatively-charged complex to neutral X and, with further hole capture, even to a $X^+$ subsequently. Moreover, it is interesting to observe an additional sharp shoulder redshifted from X with sub-barrier pumping at 900 nm, noted in Figure 1c. This is further detailed in the high-resolution spectrum of Figure 1d at 6 K. Particularly it reveals an excitonic species at an transition energy ≈ 300 $\mu$eV below the X line (Figure 1d), which matches the reported exchange interaction induced splitting between the bright and dark excitons (DX) [28]. This is detailed in Supplementary Section II. We note that the residual optical activity of the DX is attributed to mixing of the heavy hole (HH) ground state with the LH component through slightly-reduced symmetry in the QD [28, 59, 60], which results in a small in-plane polarized dipole moment and partially relieving the spin conservation.

To gain further insights on the interaction between the cavity and QD, as shown in Figure 2a, we carefully tune the X-CM coupling with fine-sweeping down to ≈ 300 mK temperature steps in our cryostat, while driving the sub-barrier excitation at 900 nm which suppresses charge fluctuation induced linewidth broadening of X from the above GaAs bandgap excitation. Figure 2b presents the X and CM peak energy as a function of temperature extracted from the spectral Lorentzian fit, with the X-CM coupling around 47 K. We witness that the slope of temperature-dependent CM energy experiences dramatic changes from 44.8 K to 46.3 K which corresponds to a detuning range 75 μeV < $\delta$ < 120 μeV, where $\delta = \omega_0 - \omega_c$ with $\omega_0$ and $\omega_c$ as the X transition and CM energies. In the negative detuning side, such energy shift appears from 47.5 K to 47.1 K (−87 μeV < $\delta$ < −83 μeV). The manner how its slopes change could manifest a trend of an avoided crossing of CM from X. Meanwhile, near zero-detuning, the CM shows a visible blue-shift crossing X with increasing temperature (data marked by blue arrow in Figure 2b). The temperature-dependent CM energy recovers its slope at far detuning. As shown in Figure 2c, the CM and X linewidths show an averaging tendency when approaching the zero-detuning temperature, albeit the CM linewidth minima does not appear at the same detuning as the X linewidth maxima. It is worth pointing out that the detuning range with CM linewidth minima corresponds to its dramatic slope changing at positive detuning. Meanwhile, the X linewidth maximum occurs at the position where the CM energy is crossing the X, with a clear blue-shift of the CM line near zero-detuning. Figure 2d further presents the ratio of integrated PL intensity of CM (and X) to the total intensity, indicating their inversion with each other as QD-CM coupling is controlled from far-detuning to resonance.

We note that the observed slope distortion of temperature-dependent CM energy appears in small detuning range (< 120 μeV) from 44 K to 46 K. Importantly, the observed avoided crossing of CM is fundamentally different from the mode pulling effect (namely, a tendency of CM energy shifting towards X near resonance), the latter of which results from the spectral overlaps of a low-$Q$ CM and the X phonon sideband [40], in the weak coupling regime. Moreover, both the avoided crossing and visible blue-shift of the CM crossing X near zero-detuning cannot be reproduced by the simple spectral overlaps between CM and X with bulk phonon dispersion as in weak coupling regime. Instead, the co-occurrence of QD-CM avoided crossing, the distinct blue-shift of the CM crossing X, and their partial out-of-synchronization linewidth averaging points to the effect of intermediate coupling mediated by phonon scattering. In this intermediate coupling regime [43], both strong and weak coupling features coexist, as studied recently theoretically. It indicates that avoided crossing can be attributed to the Rabi doublet in the canonical strong coupling regime, with the CM blue-shifted near zero-detuning as a clear signature of the phonon-mediated collective states emission in the intermediate coupling regime [43, 44].

Specifically, in this concurrent strong and weak coupling regimes, with increased phonon scattering as the bath temperature increases, each eigenstate derived by Jaynes-Cumming (JC) model experiences a different increase of the dephasing rate. The optical transition of lower and upper polaritons in the first rung of the JC ladder experiences less dephasing rate, which resembles the strong coupling feature, and causes the observed avoided crossing of CM with X in our system. Simultaneously, the higher-order eigenstates experience larger dephasing and form a collective state [43, 44], located around the center of zero-detuning, with the resultant blue shift of the coupled QD-CM emission around zero-detuning. Furthermore, we also observe an abrupt change of CM energy when approaching zero-detuning (marked by blue arrows in Figure 2b) – this is a signature of the discrete phonon modes confined in the QD structure [61]. Linewidth averaging is a canonical signature of strong coupling regime, where the CM and X linewidths collapse to the averaged value due to their photon-matter polariton nature. As to the observed partially out-of-synchronization QD-CM linewidth averaging, it indicates the mixture of the collective state and fundamental Rabi doublet [43, 44], which perturbs the exciton polariton linewidth averaging in the strong coupling regime.

From the point of view of the effective decay rate and cavity $Q$ [62], the effective CM linewidth can be renormalized by an effective cavity pumping $P_{CM}$ mainly resulting from phonon-mediated energy transfer from X to CM by $\Gamma_{CM}^{eff} = \gamma_{CM} - P_{CM}$. This resembles the CM linewidth narrowing when approaching the resonance. The reversed process applies to the effective linewidth broadening of X near the cavity resonance. We note here that our site-controlled single QD-cavity geometry safely excludes the spurious pumping terms from the parasitic QD and wetting layer in the $P_{CM}$, which commonly occurs in SK ensemble QDs. In addition, for the same cavity-QD with occasional CM blue-shift as a result of the PhC cavity surface state changes, the avoided crossing can occur at lower temperatures such as 25 K. Here the cavity $Q$ factor is up to ≈ 9,270 (135 ± 11 µeV), resulting in *solely* the QD-CM linewidth averaging on resonance (not the concurrent strong-weak coupling regimes), in the canonical strong coupling regime. This is detailed in Supplementary Section III.

**Span of degree-of-linear polarizations in the concurrent strong-weak intermediate coupling regime**

To further investigate phonon-assisted coupling in the intermediate coupling regime, we performed the polarization-resolved detuning-dependent µPL on the same cavity-QD with controlled polarizers and bath temperature, as illustrated in Figure 3a. Here, the degree-of-linear polarization (DOLP) [36, 49], defined as $\text{DOLP} = \frac{I_V - I_H}{I_V + I_H}$ where $I_V$ and $I_H$ are the vertical- and horizontal-polarized µPL components align with TE and TM cavity modes respectively, is examined. The DOLP of different exciton species at various CM detuning serves as a probe to understand the pure dephasing mediated QD-CM coupling. Figure 3b summarizes the DOLP of X and excited LH at the detuning range covering the range in Figure 2. It shows that the X and LH exciton species are co-polarized (positive DOLP) with CM near resonance. It gradually loses the CM-like polarization (DOLP from +0.6 to 0) when detuned from resonance to ≈ ± 1.2 meV detuning where the subsystem approaches the zero-DOLP (grey bar region).

Importantly, the detuning range for the pronounced co-polarization (DOLP > 0.5) corresponds to the detuning range (≈-110 µeV to +170 µeV) where dramatic slope distortion appears. At such detuning, the vertical-polarized components of QD excitation can be transferred into the cavity decay channel as a result of the phonon-assisted coupling between the CM and QD [36, 38], leading to a relative depletion of the vertical-polarized QD µPL energies. In succession, such depletion subsequently enhances the vertical-polarized decay channel of QD itself by the Purcell enhanced cavity photon channeling to the QD µPL energy, resulting in the significant co-polarized QD µPL of DOLP > 0.5 with the CM. Because the cavity loss rate is generally much larger than the QD radiative decay rate, we observe the overall co-polarization effect at detunings where QD and CM sufficiently interact. And the significant co-polarization is in roughly the same detuning range where the dramatic slope changes appear in Figure 2b, indicating the important contribution of the phonon-mediated Purcell enhancement in our observed concurrent strong-weak coupling regime. The fact that the CM modifies the QD polarization only in relatively small detuning range strongly suggest the absence of far-off-resonant coupling induced by wetting layers or background emissions.

We also observed a negative DOLP at much larger detuning range, which manifest an overall S-shape DOLP curve [36, 51], unique to these site-controlled QD-cavity system. The polarization-resolved measurements also enable the determination of the fine-structure splitting of X with orthogonal polarization, complying with the spin conservation selection rules. The inset of Figure 3b shows this polarization dependence, with a resulting extracted X fine-structure splitting of ≈ 30 μeV.

**Anti-bunching, bunching, and collective state photon statistics in the concurrent strong-weak intermediate coupling regime**

To study the radiative recombination dynamics of our coupled QD-cavity system, we measure the detuning-dependent decay time of $X^-$ by TRPL, as shown in Figure 4a. When $X^-$ is near resonance with the cavity under 160 μW excitation, Purcell enhancement results in a 1.2 ns decay time, extracted by the single-exponential decay convolved with an ≈ 500 ps instrument response function. To examine the μPL decay of the $X^-$ far-detuned from resonance, the excitation power is doubled to obtain adequate signal-to-noise ratio. The μPL rise time is clearly time-delayed, which arises from the finite p-state occupation under higher-power excitation [63]. The extracted $X^-$ off-resonance decay time increases up to 3.0 ns. We subsequently derive the Purcell factor [63] via: $\frac{\tau_0}{\tau_{X^-}(\delta)} = \frac{F_P(1+2Q\gamma_d)}{8\left(\frac{\delta}{\kappa_{CM}}\right)^2 + 2(1+2Q\gamma_d)^2} f^2 + \frac{\tau_0}{\tau_{leak}}$, where $\tau_{X^-}(\delta) = 1.2$ ns is the $X^-$ near-resonance lifetime, $\tau_{leak} = 3.0$ ns is the off-resonance $X^-$ lifetime, $\tau_0 = 1$ ns is the typical bulk exciton lifetime, $\delta = 130$ μeV is the near-resonance detuning, $Q = 5,500$ is the near-resonance $Q$ factor, $\kappa_{CM} = 227$ μeV is the near-resonance cavity linewidth, and $\gamma_d = 9.5\times10^{-5}$ is inverse $X^-$ quality factor. $f$ is a dimensionless constant, which depends on the spatial alignment between the site-controlled QD and the cavity field maxima, and the orientation matching between the QD dipole and cavity field. With $g = \sqrt{\frac{\Delta E^2}{4\hbar^2} + \frac{(\kappa_{CM}-\gamma_{X^-})^2}{16}}$ [64], we estimate the Rabi coupling strength ($g$) of around 50 μeV. The corresponding Purcell factor $F_P$ is estimated to be between 2.7 to 10.8, with the associated reasonable range of $f$ between 100% to 50%.

Photon statistics of the neutral exciton X coupled to the cavity is further investigated by measuring second-order correlation function $g^2(\tau)$. As shown in Figure 4b, at low excitation of 150 nW, we obtain a clear anti-bunching of 0.2, which supports that the coupled X-CM is at the single-photon single-exciton level. The time-bin width is chosen at ≈ 2 ns, as a consequence of trading off the temporal resolution versus the signal-to-noise ratio. We fit the anti-bunching by using a time-bin-convolved second-order autocorrelation function of two-level system $g^2(\tau) = (1 - e^{-|\tau|/\tau_d})*boxcar(\tau)$, where *boxcar*($\tau$) is the boxcar function accounting for the 2 ns time bin, and $\tau_d$ = 1.2 ns is the on-resonance decay time of X. With such fitting parameters, the fit well reproduces the observed anti-bunching signature of the coupled X-CM emission. The non-zero antibunching can be due to limited cavity feeding by X⁻ or LH, leading to an occasional uncorrelated photon. The fluctuation of the $g^2(\tau)$ baseline is attributed to a limited counting rate of the HBT setup. With doubling of the pump power to 300 nW, we witness a pronounced bunching, implying a higher conditional probability of a finding a pair of photons in our coupled QD-cavity system. The fit for bunching uses $g^2(\tau) = (1 + Ae^{-|\tau|/\tau_d})*boxcar(\tau)$, where 1+*A* is the amplitude of photon bunching. It is worth pointing out that such super-Poisson statistics has been observed in single self-assembled SK growth QD coupled with a PhC microcavity. It is postulated that the cascaded cavity photon emission can arise from an excitonic continuum induced by intermixing QD and wetting layer states [45, 46]. Those samples have a wetting layer where, driving the subsystem below saturation, detection of a cavity photon heralds an increase of probability finding the QD in an excited continuum, resulting in a second-photon emission event. However, the *absence* of the wetting layer and parasite QDs in our single pyramidal QD rules out such cascaded photon cavity feeding from the excitonic continuum or background.

Instead one can tentatively ascribe the observed bunching to a purely photon-statistical effect: during Purcell enhanced emission of the first photon in the coupled QD-CM system, the transient decrease of X means the population probability allows the subsystem to enter into the sub-Poisson statistical regime, resulting in a temporal-correlated second single-photon emission [65]. Here, since the dynamics of the X repopulation determines the difference in mean population probability between the zero and longer delays [66], a proper moderate excitation aids in the visibility of this bunching feature. One can also consider that an increased pump power induces the elevated photon number accumulation in the cavity Fock states. The resulting increased collective state [67, 68] emission dominates the photon statistics as a bunching signature. Our observed tunability of the photon number statistics indicates the collective state due to phonon-mediated coupling in the intermediate coupling regime, shedding light on the concurrent strong-weak coupling regimes form different rungs of the Jaynes-Cummings ladder in our singular site-controlled cavity-QD system.

In sum, we investigate the intrinsic dephasing-mediated intermediate QD-CM coupling quantum electrodynamics based on our unique singular site-controlled pyramidal InGaAs/GaAs QDs – L3 PhC cavities platform. The detuning-dependent µPL spectrum reveals co-occurrence of QD-CM avoided crossing, which is a signature of Rabi doublet of the strong coupled system, and the blue shifting of coupled QD-CM energy near zero-detuning due to the formation of collective states mediated by phonon assisted coupling, and their partial out-of-synchronization linewidth-narrowing linked to their mixed behavior. Further polarization-resolved µPL reveals the important contribution of the intermediate coupling from phonon-mediated coupling. Furthermore, we reveal the dynamics and anti-bunching/bunching photon statistical signatures of the concurrent strong-weak intermediate coupled system at zero-detuning. This is supported by our measurements on the exciton-photon complexes and their optical features strongly confined in our single site-controlled QD. This study suggests our device has potential for new and subtle cavity quantum electrodynamical phenomena, cavity-enhanced indistinguishable single photon generation, and cluster state generation via the exciton-photon complexes for quantum networks.

**Acknowledgement**

The authors thank helpful discussions with Abhinav Kumar Vinod, James F. McMillan, Yujin Cho, Kai-Chi Chang, Jin Ho Kang, Justin Caram, and Baolai Liang from UCLA and technical help from Alexey Lyasota and Bruno Rigal from EPFL for the sample fabrication. J.H., W.L., and C.W.W. acknowledge support from the National Science Foundation (1741707, 1936375, and 1919355). W.L. also acknowledges support from the Swiss National Science Foundation under project 187963. W.L and J.H. led the project and performed the measurements with data analysis. A.M., B. D., and A. R. grew the samples and performed the photonic crystal microcavity fabrication. C.W.W. and E.K. aided in the project. W.L., J.H, and C.W.W. wrote the manuscript, with contributions from all authors.

**References**

[1] Ekert, A. K. Quantum Cryptography Based on Bellâs Theorem. *Phys. Rev. Lett.* **67**, 661–663 (1991).

[2] Dada, A. C., Leach, J., Buller, G. S., Padgett, M. J. & Andersson, E. Experimental High-Dimensional Two-Photon Entanglement and Violations of Generalized Bell Inequalities.

Nat. Phys. **7**, 677–680 (2011).

[3]  Gao, W. B., Fallahi, P., Togan, E., Miguel-Sanchez, J. & Imamoglu, A. Observation of Entanglement between a Quantum Dot Spin and a Single Photon. *Nature* **491**, 426–430 (2012).

[4]  Weber, J. H., Kambs, B., Kettler, J., Kern, S., Maisch, J., Vural, H., Jetter, M., Portalupi, S. L., Becher, C. & Michler, P. Two-Photon Interference in the Telecom C-Band after Frequency Conversion of Photons from Remote Quantum Emitters. *Nature Nanotechnology* **14**, 23–26 (2019).

[5]  Kim, J.-H., Aghaeimeibodi, S., Carolan, J., Englund, D. & Waks, E. Hybrid Integration Methods for On-Chip Quantum Photonics. *Optica* **7**, 291 (2020).

[6]  Briegel, H.-J. J., Dür, W., Cirac, J. I. & Zoller, P. Quantum Repeaters: The Role of Imperfect Local Operations in Quantum Communication. *Phys. Rev. Lett.* **81**, 5932–5935 (1998).

[7]  Reinhard, A., Volz, T., Winger, M., Badolato, A., Hennessy, K. J., Hu, E. L. & Imamoglu, A. Strongly Correlated Photons on a Chip. *Nat. Photonics* **6**, 93–96 (2011).

[8]  Kim, H., Bose, R., Shen, T. C., Solomon, G. S. & Waks, E. A Quantum Logic Gate between a Solid-State Quantum Bit and a Photon. *Nat. Photonics* **7**, 373–377 (2013).

[9]  Sun, S., Kim, H., Luo, Z., Solomon, G. S. & Waks, E. A Single-Photon Switch and Transistor Enabled by a Solid-State Quantum Memory. *Science (80-. ).* **361**, 57–60 (2018).

[10] Volz, T., Reinhard, A., Winger, M., Badolato, A., Hennessy, K. J., Hu, E. L. & Imamoğlu, A. Ultrafast All-Optical Switching by Single Photons. *Nat. Photonics* **6**, 605–609 (2012).

[11] Kako, S., Santori, C., Hoshino, K., Götzinger, S., Yamamoto, Y. & Arakawa, Y. A Gallium Nitride Single-Photon Source Operating at 200 K. *Nat. Mater.* **5**, 887–892 (2006).

[12] Yoshle, T., Scherer, A., Hendrickson, J., Khitrova, G., Gibbs, H. M., Rupper, G., Ell, C., Shchekin, O. B. & Deppe, D. G. Vacuum Rabi Splitting with a Single Quantum Dot in a Photonic Crystal Nanocavity. *Nature* **432**, 200–203 (2004).

[13] Hennessy, K., Badolato, A., Winger, M., Gerace, D., Atatüre, M., Gulde, S., Fält, S., Hu, E. L. & Imamoğlu, A. Quantum Nature of a Strongly Coupled Single Quantum Dot-Cavity System. *Nature* **445**, 896–899 (2007).

[14] Hammerer, K., Wallquist, M., Genes, C., Ludwig, M., Marquardt, F., Treutlein, P., Zoller, P., Ye, J. & Kimble, H. J. Strong Coupling of a Mechanical Oscillator and a Single Atom. *Phys. Rev. Lett.* **103**, 063005 (2009).

[15] Bose, R., Cai, T., Choudhury, K. R., Solomon, G. S. & Waks, E. All-Optical Coherent Control of Vacuum Rabi Oscillations. *Nat. Photonics* **8**, 858–864 (2014).

[16] Liu, Y.-C., Luan, X., Li, H.-K., Gong, Q., Wong, C. W. & Xiao, Y.-F. Coherent Polariton Dynamics in Coupled Highly Dissipative Cavities. *Phys. Rev. Lett.* **112**, 213602 (2014).

[17] Gao, J., Combrie, S., Liang, B., Schmitteckert, P., Lehoucq, G., Xavier, S., Xu, X., Busch, K., Huffaker, D. L., De Rossi, A. & Wong, C. W. Strongly Coupled Slow-Light Polaritons in One-Dimensional Disordered Localized States. *Sci. Rep.* **3**, 1–6 (2013).

[18] Badolato, A., Hennessy, K., Atatüre, M., Dreiser, J., Hu, E., Petroff, P. M. & Imamoğlu, A. Deterministic Coupling of Single Quantum Dots to Single Nanocavity Modes. *Science (80-. ).* **308**, 1158–1161 (2005).

[19] Muller, A., Fang, W., Lawall, J. & Solomon, G. S. Creating Polarization-Entangled Photon Pairs from a Semiconductor Quantum Dot Using the Optical Stark Effect. *Phys. Rev. Lett.* **103**, 217402 (2009).

[20] Xie, Z., Zhong, T., Shrestha, S., Xu, X., Liang, J., Gong, Y.-X., Bienfang, J. C., Restelli, A., Shapiro, J. H., Wong, F. N. C. & Wong, C. W. Harnessing High-Dimensional Hyperentanglement through a Biphoton Frequency Comb. *Nat. Photonics* **9**, 536–542 (2015).

[21] Cheng, X., Chang, K.-C., Xie, Z., Lee, Y. S., Sarihan, M. C., Kumar, A., Li, Y., Kocaman, S., Zhong, T., Yu, M., Kwong, D.-L., Shapiro, J. H., Wong, F. N. C. & Wong, C. W. An Efficient On-Chip Single-Photon SWAP Gate for Entanglement Manipulation. in *Conference on Lasers and Electro-Optics* FM2R.5 (OSA, 2020). doi:10.1364/CLEO_QELS.2020.FM2R.5.

[22] Liu, J., Su, R., Wei, Y., Yao, B., Silva, S. F. C. da, Yu, Y., Iles-Smith, J., Srinivasan, K., Rastelli, A., Li, J. & Wang, X. A Solid-State Source of Strongly Entangled Photon Pairs with High Brightness and Indistinguishability. *Nat. Nanotechnol.* **14**, 586–593 (2019).

[23] Arsenault, A. C., Clark, T. J., Von Freymann, G., Cademartiri, L., Sapienza, R., Bertolotti, J., Vekris, E., Wong, S., Kitaev, V., Manners, I., Wang, R. Z., John, S., Wiersma, D. & Ozin, G. A. From Colour Fingerprinting to the Control of Photoluminescence in Elastic Photonic Crystals. *Nat. Mater.* **5**, 179–184 (2006).

[24] Sun, S., Kim, H., Solomon, G. S. & Waks, E. A Quantum Phase Switch between a Single Solid-State Spin and a Photon. *Nat. Nanotechnol.* **11**, 539–544 (2016).

[25] Lindner, N. H. & Rudolph, T. Proposal for Pulsed On-Demand Sources of Photonic Cluster State Strings. *Phys. Rev. Lett.* **103**, 113602 (2009).

[26] Russo, A., Barnes, E. & Economou, S. E. Photonic Graph State Generation from Quantum Dots and Color Centers for Quantum Communications. *Phys. Rev. B* **98**, 085303 (2018).

[27] Gimeno-Segovia, M., Rudolph, T. & Economou, S. E. Deterministic Generation of Large-Scale Entangled Photonic Cluster State from Interacting Solid State Emitters. *Phys. Rev. Lett.* **123**, 070501 (2019).

[28] Schwartz, I., Schmidgall, E. R., Gantz, L., Cogan, D., Bordo, E., Don, Y., Zielinski, M. & Gershoni, D. Deterministic Writing and Control of the Dark Exciton Spin Using Single Short Optical Pulses. *Phys. Rev. X* **5**, 011009 (2015).

[29] Schwartz, I., Cogan, D., Schmidgall, E. R., Don, Y., Gantz, L., Kenneth, O., Lindner, N. H. & Gershoni, D. Deterministic Generation of a Cluster State of Entangled Photons. *Science (80-. ).* **354**, 434–437 (2016).

[30] Layden, D., Chen, M. & Cappellaro, P. Efficient Quantum Error Correction of Dephasing Induced by a Common Fluctuator. *Phys. Rev. Lett.* **124**, 020504 (2020).

[31] Schöll, E., Hanschke, L., Schweickert, L., Zeuner, K. D., Reindl, M., Covre da Silva, S. F., Lettner, T., Trotta, R., Finley, J. J., Müller, K., Rastelli, A., Zwiller, V. & Jöns, K. D. Resonance Fluorescence of GaAs Quantum Dots with Near-Unity Photon Indistinguishability. *Nano Lett.* **19**, 2404–2410 (2019).

[32] Nawrath, C., Olbrich, F., Paul, M., Portalupi, S. L., Jetter, M. & Michler, P. Coherence and Indistinguishability of Highly Pure Single Photons from Non-Resonantly and Resonantly Excited Telecom C-Band Quantum Dots. *Appl. Phys. Lett.* **115**, 023103 (2019).

[33] Auffèves, A., Gerace, D., Gérard, J.-M., Santos, M. F., Andreani, L. C. & Poizat, J.-P. Controlling the Dynamics of a Coupled Atom-Cavity System by Pure Dephasing. *Phys. Rev. B* **81**, 245419 (2010).

[34] Lodahl, P., Van Driel, A. F., Nikolaev, I. S., Irman, A., Overgaag, K., Vanmaekelbergh, D. & Vos, W. L. Controlling the Dynamics of Spontaneous Emission from Quantum Dots by Photonic Crystals. *Nature* **430**, 654–657 (2004).

[35] Bayer, M. & Forchel, A. Temperature Dependence of the Exciton Homogeneous Linewidth in (Formula Presented) Self-Assembled Quantum Dots. *Phys. Rev. B - Condens. Matter Mater. Phys.* **65**, 1–4 (2002).

[36] Jarlov, C., Wodey, Lyasota, A., Calic, M., Gallo, P., Dwir, B., Rudra, A. & Kapon, E. Effect of Pure Dephasing and Phonon Scattering on the Coupling of Semiconductor Quantum Dots to Optical Cavities. *Phys. Rev. Lett.* **117**, 076801 (2016).

[37] Naesby, A., Suhr, T., Kristensen, P. T. & Mørk, J. Influence of Pure Dephasing on Emission Spectra from Single Photon Sources. *Phys. Rev. A - At. Mol. Opt. Phys.* **78**, 045802 (2008).

[38] Auffèves, A., Gérard, J. M. & Poizat, J. P. Pure Emitter Dephasing: A Resource for Advanced Solid-State Single-Photon Sources. *Phys. Rev. A - At. Mol. Opt. Phys.* **79**, 053838 (2009).

[39] Cui, G. & Raymer, M. G. Emission Spectra and Quantum Efficiency of Single-Photon Sources in the Cavity-QED Strong-Coupling Regime. *Phys. Rev. A - At. Mol. Opt. Phys.* **73**, (2006).

[40] Valente, D., Suffczyński, J. S., Jakubczyk, T., Dousse, A., Lemaître, A. L., Sagnes, I., Lanco, L. L., Voisin, P., Auffèves, A. & Senellart, P. Frequency Cavity Pulling Induced by a Single Semiconductor Quantum Dot. *RAPID Commun. Phys. Rev. B* **89**, 41302 (2014).

[41] Giesz, V. *Cavity-enhanced Photon-Photon Interactions With Bright Quantum Dot Sources*. https://tel.archives-ouvertes.fr/tel-01272948.

[42] Tawara, T., Kamada, H., Tanabe, T., Sogawa, T., Okamoto, H., Yao, P., Pathak, P. K. & Hughes, S. Cavity-QED Assisted Attraction between a Cavity Mode and an Exciton Mode in a Planar Photonic-Crystal Cavity. *Opt. Express* **18**, 2719 (2010).

[43] Echeverri-Arteaga, S., Vinck-Posada, H. & Gómez, E. A. The Strange Attraction Phenomenon in CQED: The Intermediate Quantum Coupling Regime. *Optik (Stuttg).* **183**, 389–394 (2019).

[44] Echeverri-Arteaga, S., Vinck-Posada, H. & Gómez, E. A. Explanation of the Quantum Phenomenon of Off-Resonant Cavity-Mode Emission. *Phys. Rev. A* **97**, 43815 (2018).

[45] Winger, M., Volz, T., Tarel, G., Portolan, S., Badolato, A., Hennessy, K. J., Hu, E. L., Beveratos, A., Finley, J., Savona, V. & Imamoğlu, A. Explanation of Photon Correlations in the Far-off-Resonance Optical Emission from a Quantum-Dot-Cavity System. *Phys. Rev. Lett.* **103**, (2009).

[46] Toda, Y., Moriwaki, O., Nishioka, M. & Arakawa, Y. Efficient Carrier Relaxation Mechanism in In Gaas/Gaas Self-Assembled Quantum Dots Based on the Existence of Continuum States. *Phys. Rev. Lett.* **82**, 4114–4117 (1999).

[47] Vasanelli, A., Ferreira, R. & Bastard, G. Continuous Absorption Background and Decoherence in Quantum Dots. *Phys. Rev. Lett.* **89**, 216804 (2002).

[48] Gallo, P., Felici, M., Dwir, B., Atlasov, K. A., Karlsson, K. F., Rudra, A., Mohan, A., Biasiol, G., Sorba, L. & Kapon, E. Integration of Site-Controlled Pyramidal Quantum Dots and Photonic Crystal Membrane Cavities. *Appl. Phys. Lett.* **92**, 263101 (2008).

[49] Calic, M., Gallo, P., Felici, M., Atlasov, K. A., Dwir, B., Rudra, A., Biasiol, G., Sorba, L., Tarel, G., Savona, V. & Kapon, E. Phonon-Mediated Coupling of InGaAs/GaAs Quantum-Dot Excitons to Photonic Crystal Cavities. *Phys. Rev. Lett.* **106**, 227402 (2011).

[50] Jarlov, C. Cavity Quantum Electrodynamics with Systems of Site-Controlled Quantum Dots and Photonic Crystal Cavities. *PhD thesis, EPFL* **7039**, (2016).

[51] Ćalić, M. Cavity Quantum Electrodynamics with Site-Controlled Pyramidal Quantum Dots in Photonic Crystal Cavities. *PhD thesis, EPFL* **5957**, (2013).

[52] Jarlov, C., Gallo, P., Calic, M., Dwir, B., Rudra, A. & Kapon, E. Bound and Anti-Bound Biexciton in Site-Controlled Pyramidal GaInAs/GaAs Quantum Dots. *Appl. Phys. Lett.* **101**, 191101 (2012).

[53] Cade, N. I., Gotoh, H., Kamada, H., Nakano, H. & Okamoto, H. Fine Structure and Magneto-Optics of Exciton, Trion, and Charged Biexciton States in Single InAs Quantum Dots Emitting at 1.3 μm. *Phys. Rev. B* **73**, 115322 (2006).

[54] Muljarov, E. A. & Zimmermann, R. Dephasing in Quantum Dots: Quadratic Coupling to Acoustic Phonons. *Phys. Rev. Lett.* **93**, 237401 (2004).

[55] Ware, M. E., Stinaff, E. A., Gammon, D., Doty, M. F., Bracker, A. S., Gershoni, D., Korenev, V. L., Bădescu, Ş. C., Lyanda-Geller, Y. & Reinecke, T. L. Polarized Fine Structure in the Photoluminescence Excitation Spectrum of a Negatively Charged Quantum Dot. *Phys. Rev. Lett.* **95**, 177403 (2005).

[56] Karlsson, K. F., Oberli, D. Y., Dupertuis, M. A., Troncale, V., Byszewski, M., Pelucchi, E., Rudra, A., Holtz, P. O. & Kapon, E. Spectral Signatures of High-Symmetry Quantum Dots and Effects of Symmetry Breaking. *New J. Phys.* **17**, 103017 (2015).

[57] Johnsson, M., Góngora, D. R., Martinez-Pastor, J. P., Volz, T., Seravalli, L., Trevisi, G., Frigeri, P. & Muñoz-Matutano, G. Ultrafast Carrier Redistribution in Single InAs Quantum Dots Mediated by Wetting-Layer Dynamics. *Phys. Rev. Appl.* **11**, 54043 (2019).

[58] Baier, M. H., Malko, A., Pelucchi, E., Oberli, D. Y. & Kapon, E. Quantum-Dot Exciton

Dynamics Probed by Photon-Correlation Spectroscopy. *Phys. Rev. B* **73**, 205321 (2006).

[59] Dupertuis, M. A., Karlsson, K. F., Oberli, D. Y., Pelucchi, E., Rudra, A., Holtz, P. O. & Kapon, E. Symmetries and the Polarized Optical Spectra of Exciton Complexes in Quantum Dots. *Phys. Rev. Lett.* **107**, 127403 (2011).

[60] Huber, D., Lehner, B. U., Csontosová, D., Reindl, M., Schuler, S., Covre Da Silva, S. F., Klenovský, P. & Rastelli, A. Single-Particle-Picture Breakdown in Laterally Weakly Confining GaAs Quantum Dots. *Phys. Rev. B* **100**, 235425 (2019).

[61] Fainstein, A., Lanzillotti-Kimura, N. D., Jusserand, B. & Perrin, B. Strong Optical-Mechanical Coupling in a Vertical GaAs/AlAs Microcavity for Subterahertz Phonons and Near-Infrared Light. *Phys. Rev. Lett.* **110**, 037403 (2013).

[62] Laussy, F. P., del Valle, E. & Tejedor, C. Luminescence Spectra of Quantum Dots in Microcavities. I. Bosons. *Phys. Rev. B* **79**, 235325 (2009).

[63] Jarlov, C., Lyasota, A., Ferrier, L., Gallo, P., Dwir, B., Rudra, A. & Kapon, E. Exciton Dynamics in a Site-Controlled Quantum Dot Coupled to a Photonic Crystal Cavity. *Appl. Phys. Lett.* **107**, 191101 (2015).

[64] Kim, H., Sridharan, D., Shen, T. C., Solomon, G. S. & Waks, E. Strong Coupling between Two Quantum Dots and a Photonic Crystal Cavity Using Magnetic Field Tuning. *Opt. Express* **19**, 2589 (2011).

[65] Regelman, D. V., Mizrahi, U., Gershoni, D., Ehrenfreund, E., Schoenfeld, W. V. & Petroff, P. M. Semiconductor Quantum Dot: A Quantum Light Source of Multicolor Photons with Tunable Statistics. *Phys. Rev. Lett.* **87**, 257401-1-257401–4 (2001).

[66] Kuroda, T., Belhadj, T., Abbarchi, M., Mastrandrea, C., Gurioli, M., Mano, T., Ikeda, N., Sugimoto, Y., Asakawa, K., Koguchi, N., Sakoda, K., Urbaszek, B., Amand, T. & Marie, X. Bunching Visibility for Correlated Photons from Single GaAs Quantum Dots. *Phys. Rev. B* **79**, 035330 (2009).

[67] Echeverri-Arteaga, S., Vinck-Posada, H., Villas-Bôas, J. M. & Gómez, E. A. Pure Dephasing vs. Phonon Mediated off-Resonant Coupling in a Quantum-Dot-Cavity System. *Opt. Commun.* **460**, 125115 (2020).

[68] Dubin, F., Russo, C., Barros, H. G., Stute, A., Becher, C., Schmidt, P. O. & Blatt, R. Quantum to Classical Transition in a Single-Ion Laser. *Nat. Phys.* **6**, 350–353 (2010).

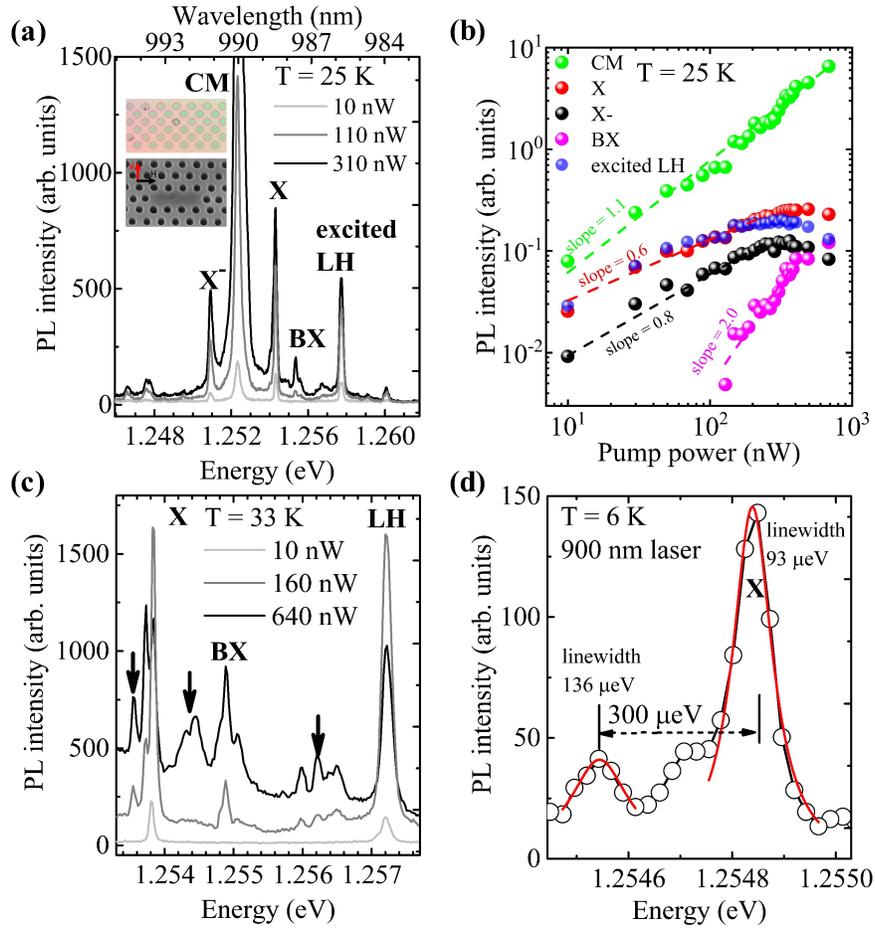

**Figure 1 | Exciton complexes and dark excitons in singular site-controlled InGaAs cavity quantum electrodynamics. a,** μPL spectrum of a singular pyramidal InGaAs QD-L3 PhC cavity system at 25 K with 532 nm non-resonant pumping at 10 nW, 110 nW, and 310 nW. **b,** Integrated μPL intensity of different QD exciton species and cavity mode as a function of pump power at 25 K. **c,** A zoom-in of the appeared ultrasharp exciton-complex species marked by arrows at 33 K (near the X and BX emission) with 532 nm non-resonant pumping at 10 nW, 160 nW, and 640 nW. **d,** Example zoom-in of the sharp peak 300 μeV below X emission, at 6 K with 900 nm pumping.

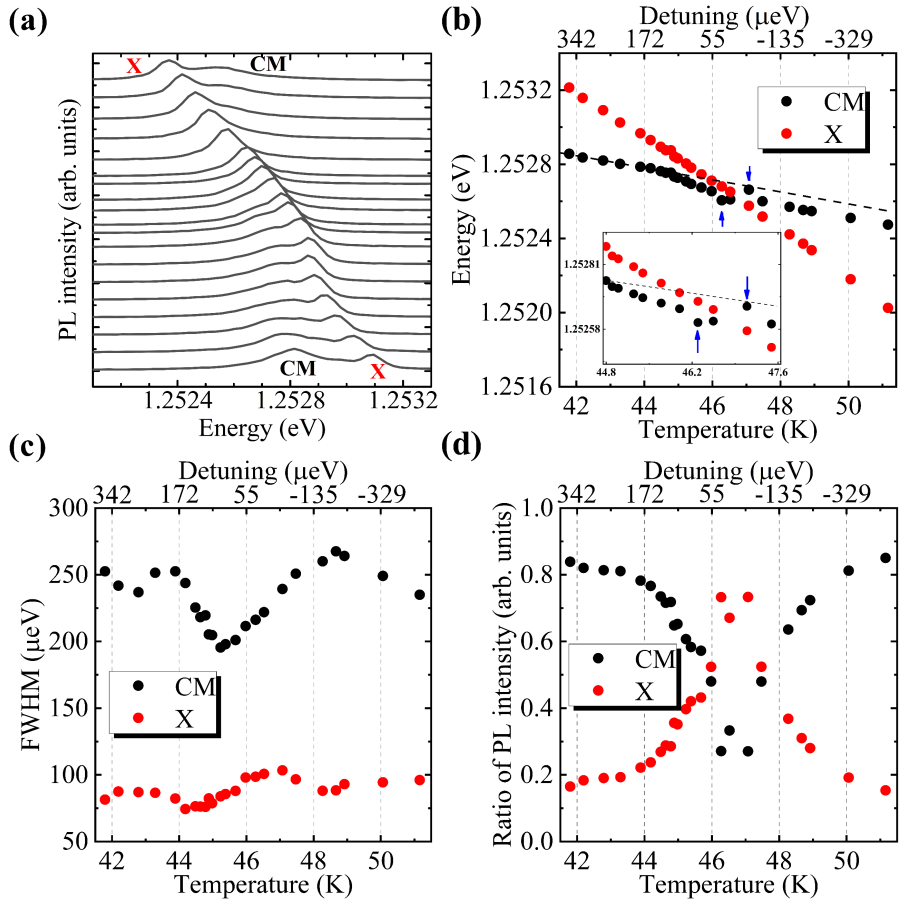

**Figure 2 | Intermediate coupling regime probed by stepped temperature dependent μPL. a,** Temperature dependent μPL traces of CM and X emission. From bottom to top: increasing temperature. **b,** Peak energy of CM and X emission as a function of temperature extracted by spectral Lorentzian fitting of a. The black dash line indicates the slopes of CM at large X-CM detuning. Inset: zoom-in of the mode crossing. **c,** Linewidth of CM and X as a function of temperature. **d,** Ratio of the integrated CM and X emission intensity as a function of temperature. Measurements are with 900 nm pulsed excitation.

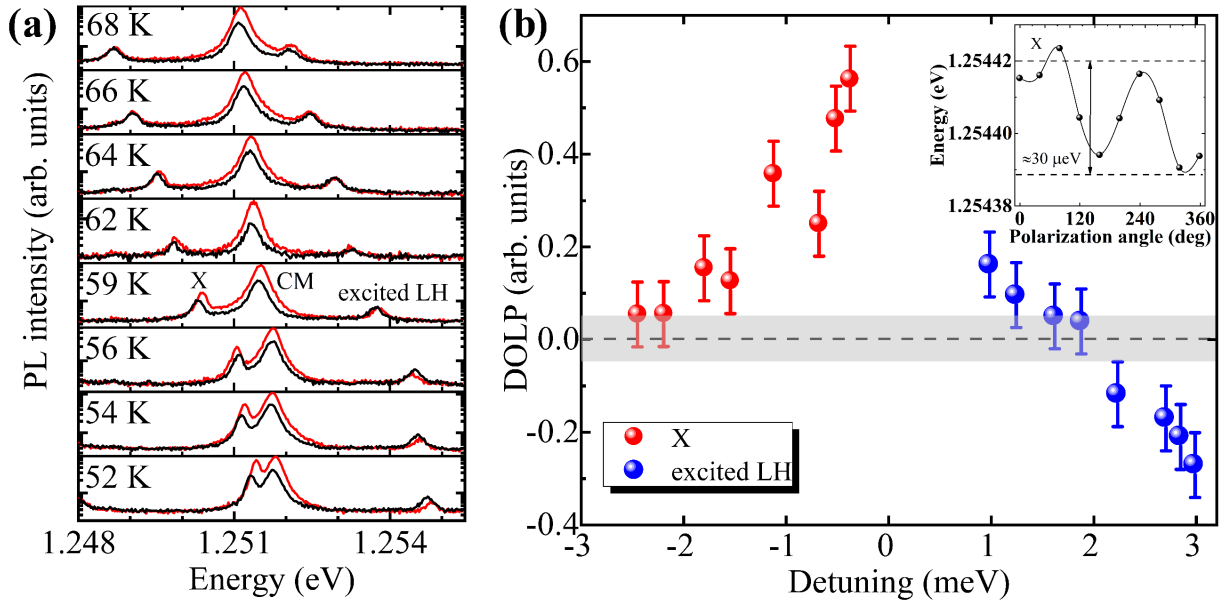

**Figure 3 | Phonon-assisted coupling examined by the degree of linear polarization (DOLP).** **a,** μPL spectrum resolved in H (black) and V (red) polarizations in the QD-CM detuning range from -3 meV to 3 meV. **b,** DOLP of X and LH as a function of their detuning with respect to CM extracted from a. The black dash line marks the zero-DOLP line, and the grey bar region is the typical DOLP values for bare QDs without the presence of a cavity. Inset: X emission energy as a function of the half-wave plate fast-axis angle, indicative of the fine-structure splitting. Detuning is defined as the energy difference with respect to CM. Measurements are with 532 nm CW excitation.

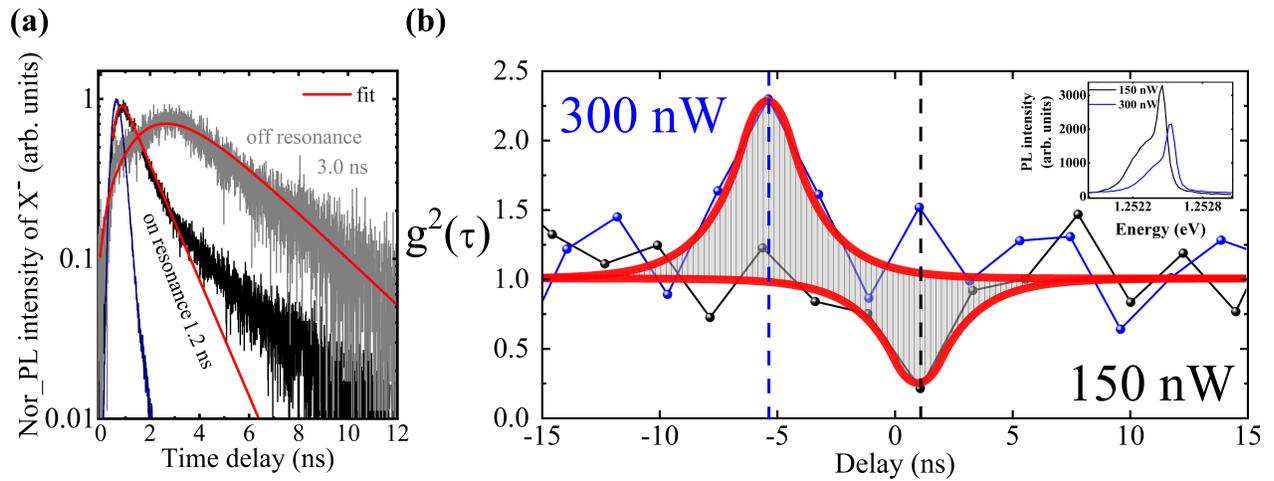

**Figure 4 | Exciton dynamics and photon statistics of the site-controlled QD-cavity subsystem.
a,** Time-resolved μPL when X- is off- and on-resonance with the CM. System instrument response is represented by the dark blue curve. The red curves are the single exponential fits. A 900 nm pulse laser is used for the excitation. **b,** Second-order correlation function measured when X is near resonance with the CM under 532 nm excitation at 150 nW (black) and 300 nW (blue). In the case of anti-bunching, the red curve is fit by a time-bin-convolved second-order autocorrelation function of two-level system $g^2(\tau) = (1 - e^{-|\tau|/\tau_d})*boxcar(\tau)$, where $boxcar(\tau)$ is the boxcar function accounting for the 2 ns time bin, and $\tau_d$ = 1.2 ns is the near-resonance decay time of X. On the other hand, the fit for bunching uses $g^2(\tau) = (1 + Ae^{-|\tau|/\tau_d})*boxcar(\tau)$, where 1+A is the amplitude of photon bunching. The black (blue) dash line indicates the delay that antibunching (bunching) occurs. Inset: μPL spectrum of near resonant X-CM at 150 nW and 300 nW.

# Supplementary Information

# Exciton-photon complexes and dynamics in the concurrent strong-weak coupling regime of singular site-controlled cavity quantum electrodynamics

Jiahui Huang[1,†], Wei Liu[1,†,*], Murat Can Sarihan[1], Xiang Cheng[1], Alessio Miranda[2], Benjamin Dwir[2], Alok Rudra[2], Eli Kapon[2], Chee Wei Wong[1,*]

[1] Mesoscopic Optics and Quantum Electronics Laboratory, University of California, Los Angeles, CA 90095, USA

[2] Institute of Physics, École Polytechnique Fédérale de Lausanne, Lausanne, Switzerland

* Correspondence: weiliu01@ucla.edu; cheewei.wong@ucla.edu

† These authors contributed equally

This Supplementary Information is organized as follows. In **Section I**, we introduce measurement schematic, describe the sample structure, and model the photonic crystal (PhC) structures. In **Section II**, we describe the power dependence of the dark exciton and its high-resolution fine structure, along with measurements in another QD-cavity sample for iterated confirmation. In **Section III**, we discuss an occasional red shift of cavity mode (CM) likely related to the variation of surface condition of PhC induced by ambient condition, which allows coupling between exciton and CM at 41K and 25 K. In **Section IV**, we describe the fundamental of QD-cavity concurrent strong-weak intermediate coupling and their governing dephasing mechanism in the Lindblad master equation formalism.

## I. Measurement schematic and sample structure

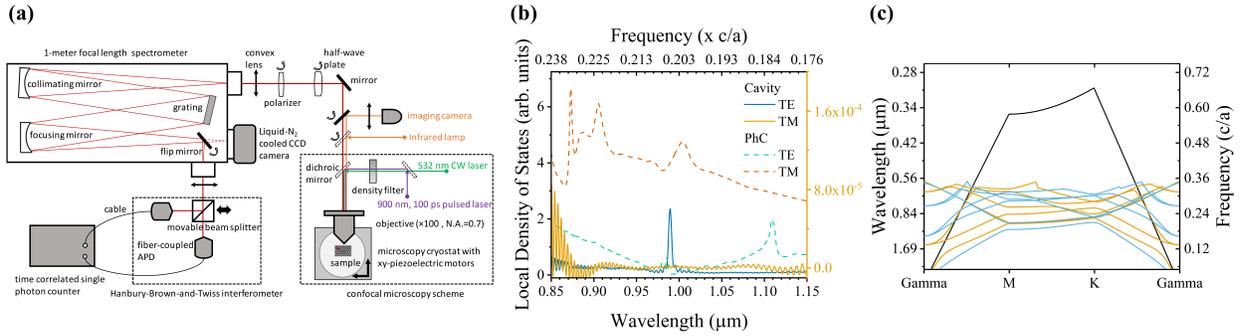

**Figure S1** | **a,** Schematic of optical characterization setup, which enables high spectral-resolved and polarization-resolved µPL, time-resolved PL, and second-order autocorrelation ($g^2$) measurements. **b,** Simulated local density of states. **c,** Energy dispersion of the *L*3 photonic crystal cavity. The right axis only corresponds to the cavity TM mode (yellow solid line) in **b**. For lattice constant *a*, air hole radius *r*, and PhC thickness *t*, the following values are used: $a$ = 202.6 nm, $r/a$ = 0.15, $t/a$ = 1.25. Refractive index of GaAs is taken to be 3.52. c is the speed of light in vacuum.

Figure S1a illustrates the schematic of our optical characterization setup, which allows high spectral-resolved and polarization-resolved µPL, time-resolved PL, and second-order autocorrelation ($g^2$) measurements.

The single site-controlled pyramidal QDs are fabricated on the (111)B-orientated GaAs substrate with electron-beam lithography (EBL) (precision within ± 5 nm) inverted pyramidal pits pattern by metalorganic vapor-phase epitaxy (MOVPE) growth of InGaAs/GaAs. The lens-shaped highly hexagonal symmetric QD is formed at the apex of highly symmetric inverted pyramid with well defined (111)A gallium terminated facets and no 2D wetting layer is formed during the growth. The size of the inverted pyramidal pit is ≈ 300 nm in base length and ≈ 250 nm in depth, and the size of QD is ≈ 20 nm in plane and ≈ 5 nm in growth direction. In some circumstance, InGaAs/GaAs quantum wires (QWR) are formed on the three wedges of the inverted pyramid during the growth. A L3 photonic crystal (PhC) structure with slab thickness ≈ 250 nm, lattice constant ≈ 200 nm, and air hole radius ≈ 30 to 42 nm is lithographically written on top of the QD pattern and all the QDs are etched away except the one at the center of the cavity. The details of QD growth and PhC cavity integration can be found in ref [S1].

With the above parameters, the numerical simulation (Figure S1b) of our PhC slab shows the band gap of ≈ 40 nm of the local density of photon (LDOS) states in the TE mode (green dashed line) and the TM mode (brown dashed line) is continuous covering all the wavelengths. The L3 PhC cavity has one isolated TE LDOS (blue solid line) at 990 nm in the photonic band gap and no TM LDOS (yellow solid line) is found in cavity from the simulation. Figure S1c shows the photon energy dispersion from Γ to M and K point.

## II. Power dependence of dark-exciton-like signature

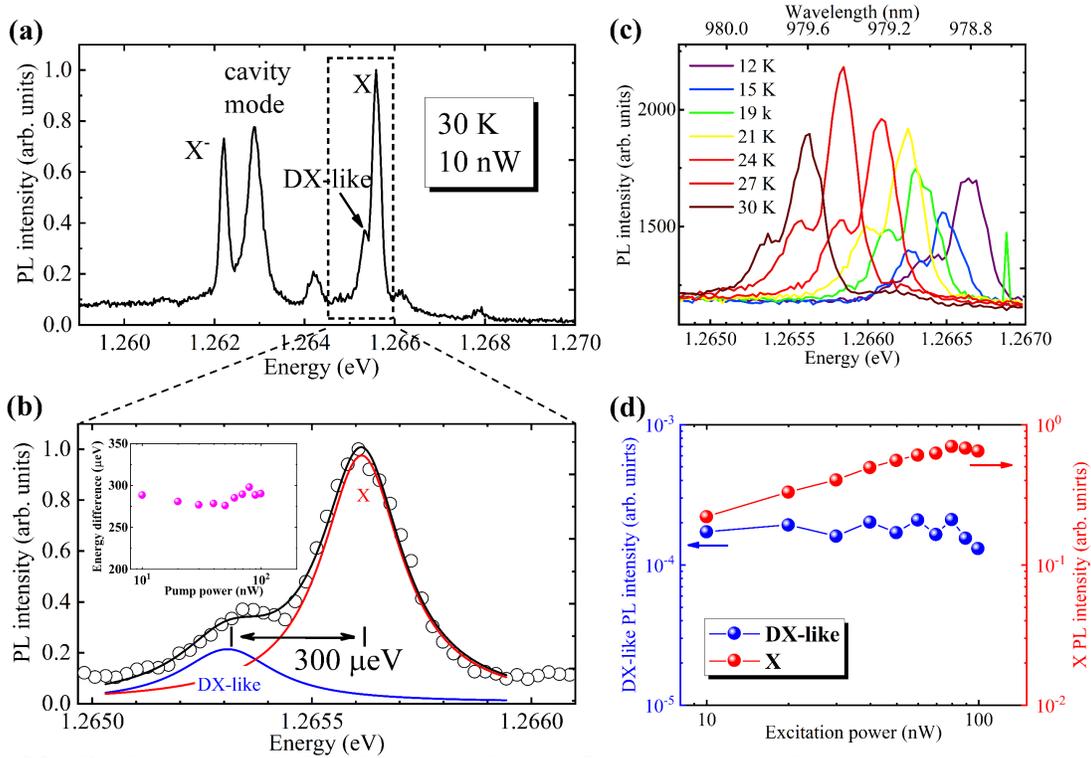

**Figure S2 | Dark exciton and its high-resolution fine structure, along with measurements in another QD-cavity sample. a,** μPL spectrum of singular pyramidal InGaAs QD-L3 PhC cavity system at 30 K with 532 nm non-resonant pumping at 10 nW. **b,** Zoom-in of the sharp peak (DX-like) approximately 300 μeV below X. Inset: energy separation between DX-like and X emission as a function of pump power. **c,** DX-like and X spectra at different temperatures. **d,** Integrated μPL intensities of DX-like and X emission as a function of pump power. All measurements are performed by using the 532 nm CW laser.

Comparing to bright exciton with antiparallel electron and hole pair, dark exciton (DX), consisting of a parallel electron and hole pair, normally presents a relative long lifetime and resulting optical inactivity because of spin and momentum conservation [S2]. Previous study reported the possible observation of DX μPL signature in self-assembled InGaAs QD and explained the residual optical activity from the reduced symmetry in a real QD and resulting bright and dark exciton mixing from electron and hole spin flipping process [S3, S4]. Figure S2a shows the high-resolution μPL spectra at 30 K measured on another QD-L3 cavity system on our sample with 532 nm excitation. A clear sharp emission can be observed at around 300 μeV below *X* at low pump power (10 nW), which reveals typical feature of DX. Additionally, the ≈ 300 μeV energy difference can be preserved up to 30 K temperature and 100 nW pump power. Figure S2b shows the zoom-in of DX-like emission and *X* Pump power dependence of μPL intensities for X and DX at 30 K are shown in Figure S2b. DX-like luminescence presents a maximum intensity three orders of magnitude weaker than X. Moreover, X saturates at around 100 nW but DX-like μPL intensity saturates at a power one or two orders of magnitude lower than X. At a higher pump power, the carrier accumulation rate could exceed the slow DX-like emission radiative decay rate which can leads to its early saturation [S3]. This may suggest a one or two orders of magnitude longer radiative lifetime (≈ 100 - 1000 ns) of DX-like emission than X. The observation of such signature on another QD-cavity system indicates the generality and repeatability of the measurement and suggest the possibility of implementing DX spin qubit in our system for coherent manipulation of semiconductor stationary and flying qubit, such as cluster state generation [S5, S6].

### III. Occasional CM shift due to PhC cavity surface state

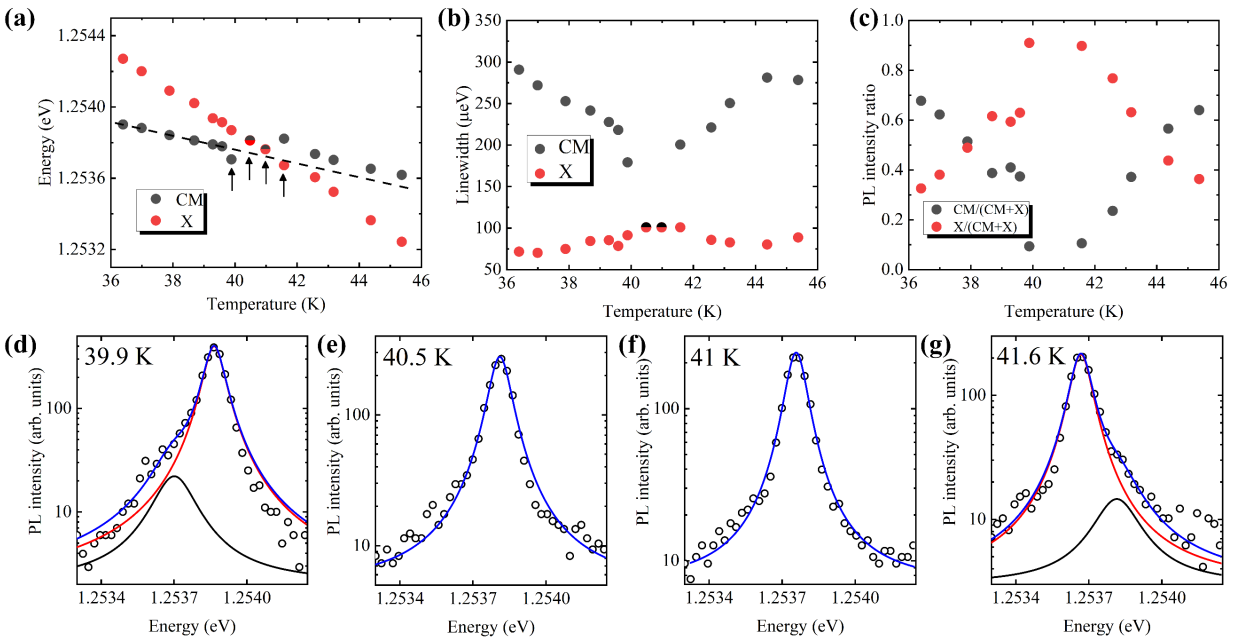

**Figure S3-1 | QD-cavity detuning by varying temperature in the case of occasional blue-shift**

**of CM due to PhC cavity surface state.** Now the *X*-CM occurs around 41 K. **a,** Emission energies. **b,** Linewidths. **c,** Ratio of PL intensity of *X* and CM emission as a function of temperature. **d-g,** Single and double Lorentzian fits of the PL spectrum for data points marked by arrows in **a**.

During our measurement, we observed occasional red shifting of CM around 1 to 3 nm. Heating the sample up to 305 K at low pressure ($10^{-6}$ Torr) for several hours can recover the original CM, which indicates the shifting can be related to the unintentional attachment of impurities from cryostat to the photonic crystal surface during the cooling process. Indeed, previous studies also reported CM shifting due to condensation of gas at the edge of air hole around the resonant cavity on the PhC slab which results in a modified slab thickness and hole diameter [S7, S8]. Figure S3-1 shows the µPL measurement for the same device in the main text after the occasional CM blue-shift, which brings the X-CM coupling to occur around 41 K. Figure S3-1 (a), (b), and (c) present the emission energy, linewidths, and ratio of PL intensity as a function of temperature. Figure S3-1 (d)-(g) are the double Lorentzian fitting of the spectrum at the data points marked by arrows in (a). The previous observed avoided crossing and blue shift are still visible suggest the repeatability of the phenomena.

The changes of PhC surface states can also bring the X-CM coupling temperature down to much lower temperature. Figure S3-2 (a) and (b) indicates that X can be close to resonance with the shifted CM at around 25 K. Our high-resolution spectrometer allows us to resolve the CM as a shoulder peak at 25 K and presents a relatively high $Q$ factor ($\approx$ 9,270) at near resonance. As suggested in the Ref. [9], the effective $Q$ of the cavity can be modified as a result of coupling to the QD. The reduction in pure dephasing of the QD should lead to an increase of $Q$ and therefore to a narrowing of the cavity linewidth. Moreover, a pronounced PL intensity raising of X and reduction of CM indicates a significant cavity photon density of state present at X emission energy near resonance (Figure S3-2(d)). Comparing with the case where X is coupled with CM at 47 K, coupling at 25 K shows an obvious linewidth exchanging indicates approaching strong coupling regime [Figure S3-2(e)] [S9, S10].

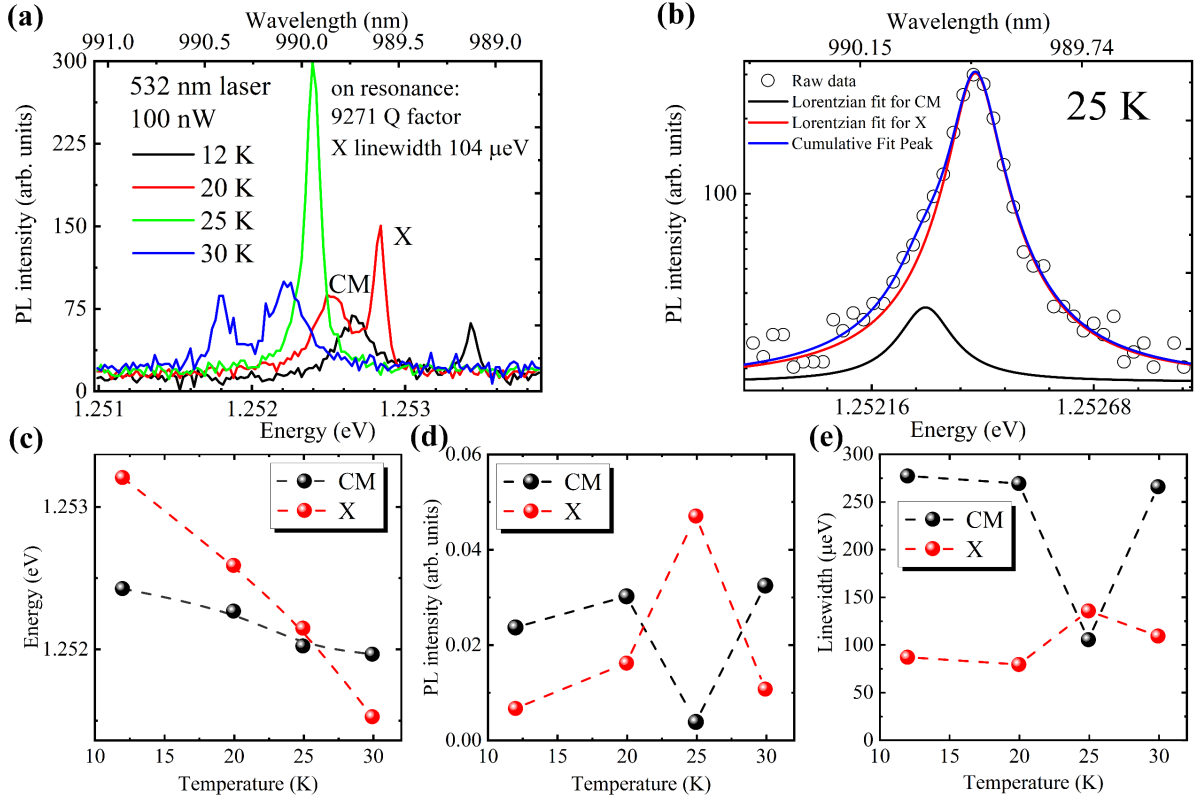

**Figure S3-2 | QD-cavity detuning by varying temperature in the case of occasional blue-shift of CM due to PhC cavity surface state. a,** μPL spectrum of another singular pyramidal InGaAs QD-L3 PhC cavity system at 30 K with 532 nm non-resonant pumping at 10 nW. **b,** Zoom-in of the X and CM spectrum and their Lorentzian fits at 25 K. **c to e,** Emission energies, integrated μPL intensities, and linewidths of X and CM emission as a function of temperature.

## IV. Description of dynamic phase change and intermediate coupling regime in the Lindblad master equation formalism

The total Hamiltonian governing the QD two-level system (TLS) and a single localized CM is given in the Jaynes-Cummings (JC) form as ($\hbar = 1$)

$$\hat{\mathcal{H}} = \omega_0 \sigma^+ \sigma^- + \omega_c a^\dagger a + g(\sigma^+ a + \sigma^- a^\dagger). \qquad (1)$$

In the above expression, $\omega_0$ and $\omega_c$ are the TLS transition energy and CM energy, respectively. $\sigma^+$ ($\sigma^-$) is the creation (annihilation) operator of the TLS and $a^\dagger$ ($a$) is the creation (annihilation) operator of the CM. TLS and CM are coupled with a coupling strength $g$.

Considering the irreversible dephasing processes in the semiconducting environment, the finite coupling of the hybrid TLS-CM system can be described in the framework of Lindblad master equation formalism written [S11, S12] as

$$\dot{\rho} = \frac{i}{\hbar}[\rho, \hat{\mathcal{H}}] + 2\kappa_C \mathcal{L}_a(\rho) + 2\gamma_X \mathcal{L}_\sigma(\rho) + 2P_X \mathcal{L}_{\sigma^+}(\rho) + 2\Gamma_d \mathcal{L}_{\sigma^- a^\dagger}(\rho), \qquad (2)$$

where the Lindblad operator with a dephasing operator $O$ is defined as

$$\mathcal{L}_O(\rho) = O\rho O^\dagger - \frac{1}{2}O^\dagger O \rho - \frac{1}{2}\rho O^\dagger O, \qquad (3)$$

With above equation, we can define the detuning between TLS and CM as $\delta = \omega_0 - \omega_c$, TLS exciton spontaneous decay rate as $2\gamma_X$, cavity loss rate as $2\kappa_C$, incoherent QD pumping rate as $2P_X$, and exciton dephasing rate as $2\Gamma_d$. Note that the dephasing term is introduced as the annihilation of QD exciton and creation of cavity photon and the reversed process is ignored which is reasonable in the case that $\kappa_C \gg \gamma_X$ [S13]. Previous experiments reveal that $\Gamma_d$ incorporates the phonon scattering rate which is a S-shape function of temperature [S14]. It shows a plateau at low temperature but features a sudden increase at increasing temperature [S14].

To gain insights into the dynamic phase transition in the cQED, previous theory proposes the substitution in the form of $\hat{K} = \hat{H} - i2\gamma_X \sigma^+ \sigma^- - i2\kappa_C a^\dagger a$ and assume $P_X = 0$ by noticing the phonon assisted coupling preserves the total number of excitations in the coupled TLS-CM system $[\mathcal{L}_{\sigma^- a^\dagger}, N_{exc}] = [\mathcal{L}_{\sigma^- a^\dagger}, \sigma^+ \sigma^- + a^\dagger a] = 0$ [S12]. The modified Lindblad master equation is now

$$\dot{\rho} = \frac{i}{\hbar}[\rho, \hat{K}] + 2\Gamma_d \mathcal{L}_{\sigma^- a^\dagger}(\rho). \qquad (4)$$

Since the Liouvillian preserves the number of excitations $N_{exc}$, it can be partitioned into four-dimensional subspaces as 4 x 4 matrices

$$\mathcal{L}^{n,n-1} = \begin{bmatrix} \kappa_C - (2n-1)\Gamma_d & -ig\sqrt{n-1} & ig\sqrt{n} & 0 \\ -ig\sqrt{n-1} & \gamma_X + Z & 0 & ig\sqrt{n} \\ ig\sqrt{n} & 0 & \kappa_C + 2\Gamma_d + Z^* - \gamma_X & -ig\sqrt{n-1} \\ 2\sqrt{n(n-1)}\Gamma_d & ig\sqrt{n} & -ig\sqrt{n-1} & \kappa_C \end{bmatrix} \qquad (5)$$

with $Z = -n\Gamma_d + 4i\delta$ and each subspace (rungs in the JC ladder), which is closely related to the emission spectrum of the system, is spanned by excitations $n$ and $m$ with eigenvalue problem $\mathcal{L}^{n,m} U^{n,m} = \lambda^{n,m} U^{n,m}$ with eigenvalues $\lambda^{n,m}$ and associated eigenfunctions $U^{n,m}$ [S12]. In the case of finite $\Gamma_d$, diagonalizing $\mathcal{L}^{n,n-1}$ produces four eigenvalues $\lambda^{n,n-1}_{\pm,\pm}$ but only two contribute to the emission spectrum of the optical transition in upper and lower polariton states of the JC rungs $n$ and $n-1$ significantly with peak energies $\omega_{-,\pm}^{n,n-1} = \text{Im}[\lambda_{-,\pm}^{n,n-1}]$ and linewidths $\Gamma_{-,\pm}^{n,n-1} = \text{Re}[\lambda_{-,\pm}^{n,n-1}]$ [S12]. With the above derivation, the measured PL spectrum in the coupled TLS-CM system will be a collective optical transition of the eigenfrequencies in all the subspaces (rungs). Furthermore, for each subspace $n$, these eigenenergies approach each other and merge to critical values at an exceptional point (EP) $\Gamma_d^{(n)} \approx 4g\sqrt{\frac{(4n_1^3+16n_1^2+10n_1+6)^{1/2}-(2n^3-3n^2+n)-\frac{2(\kappa_C-\gamma_X)}{n(n+1)}}{15n_1^2+10n_1+6}}$, where $n_1 = n(n-1)$ for $n = 2, 3, ...$ and $\Gamma_d^{(1)} = 4g - 2(\kappa_C - \gamma_X)$, which indicates a dynamical phase transition (DPT) in the system leading to a phenomenology of coexistence of strong and weak coupling regime [S12, S15].

**Supplementary References**

[S1] Ćalić, M. Cavity Quantum Electrodynamics with Site-Controlled Pyramidal Quantum Dots in Photonic Crystal Cavities. *PhD thesis, EPFL* **5957**, (2013).

[S2] Poem, E., Kodriano, Y., Tradonsky, C., Lindner, N. H., Gerardot, B. D., Petroff, P. M. & Gershoni, D. Accessing the Dark Exciton with Light. *Nat. Phys.* **6**, 993–997 (2010).

[S3] Schwartz, I., Schmidgall, E. R., Gantz, L., Cogan, D., Bordo, E., Don, Y., Zielinski, M. & Gershoni, D. Deterministic Writing and Control of the Dark Exciton Spin Using Single Short Optical Pulses. *Phys. Rev. X* **5**, 011009 (2015).

[S4] Ortner, G., Bayer, M., Larionov, A., Timofeev, V. B., Forchel, A., Lyanda-Geller, Y. B., Reinecke, T. L., Hawrylak, P., Fafard, S. & Wasilewski, Z. Fine Structure of Excitons in [Formula Presented] Coupled Quantum Dots: A Sensitive Test of Electronic Coupling. *Phys. Rev. Lett.* **90**, 4 (2003).

[S5] Gimeno-Segovia, M., Rudolph, T. & Economou, S. E. Deterministic Generation of Large-Scale Entangled Photonic Cluster State from Interacting Solid State Emitters. *Phys. Rev. Lett.* **123**, 070501 (2019).

[S6] Schwartz, I., Cogan, D., Schmidgall, E. R., Don, Y., Gantz, L., Kenneth, O., Lindner, N. H. & Gershoni, D. Deterministic Generation of a Cluster State of Entangled Photons. *Science (80-. ).* **354**, 434–437 (2016).

[S7] Mosor, S., Hendrickson, J., Richards, B. C., Sweet, J., Khitrova, G., Gibbs, H. M., Yoshie, T., Scherer, A., Shchekin, O. B. & Deppe, D. G. Scanning a Photonic Crystal Slab

Nanocavity by Condensation of Xenon. *Appl. Phys. Lett.* **87**, 1–3 (2005).

[S8]  Srinivasan, K. & Painter, O. Linear and Nonlinear Optical Spectroscopy of a Strongly Coupled Microdisk–Quantum Dot System. *Nature* **450**, 862–865 (2007).

[S9]  Auffèves, A., Gerace, D., Gérard, J.-M., Santos, M. F., Andreani, L. C. & Poizat, J.-P. Controlling the Dynamics of a Coupled Atom-Cavity System by Pure Dephasing. *Phys. Rev. B* **81**, 245419 (2010).

[S10] Borri, P., Langbein, W., Schneider, S., Woggon, U., Sellin, R. L., Ouyang, D. & Bimberg, D. Ultralong Dephasing Time in Ingaas Quantum Dots. *Phys. Rev. Lett.* **87**, (2001).

[S11] Cheng, X., Ye, H. & Yu, Z. Unconventional Photon Blockade in a Photonic Molecule Containing a Quantum Dot. *Superlattices Microstruct.* **105**, 81–89 (2017).

[S12] Echeverri-Arteaga, S., Vinck-Posada, H. & Gómez, E. A. Explanation of the Quantum Phenomenon of Off-Resonant Cavity-Mode Emission. *Phys. Rev. A* **97**, 43815 (2018).

[S13] Hohenester, U. Cavity Quantum Electrodynamics with Semiconductor Quantum Dots: Role of Phonon-Assisted Cavity Feeding. *Phys. Rev. B - Condens. Matter Mater. Phys.* **81**, (2010).

[S14] Laucht, A., Hauke, N., Villas-Bôas, J. M., Hofbauer, F., Böhm, G., Kaniber, M. & Finley, J. J. Dephasing of Exciton Polaritons in Photoexcited InGaAs Quantum Dots in GaAs Nanocavities. *Phys. Rev. Lett.* **103**, 087405 (2009).

[S15] Echeverri-Arteaga, S., Vinck-Posada, H. & Gómez, E. A. The Strange Attraction Phenomenon in CQED: The Intermediate Quantum Coupling Regime. *Optik (Stuttg).* **183**, 389–394 (2019).